\documentclass[useAMS,usenatbib]{mn2e}
\usepackage{graphicx}
\usepackage{aas_macros}
\usepackage{amssymb}

\title[Unveiling the redback nature of the low-mass X-ray binary 
XSS\,J1227.0-4859 through optical observations]
{Unveiling the redback nature of the low-mass X-ray binary XSS\,J1227.0-4859
through optical observations\thanks{Based on observations made at ESO and SALT telescopes}}
\author[D. de Martino et al.]{D.~de Martino,$^{1}$ \thanks{E-mail: demartino@oacn.inaf.it}
J.~Casares$^{2}$, 
E.~Mason$^{3}$, 
D.A.H.~Buckley$^{4}$, 
M.M.~Kotze$^{5}$,
\newauthor
J.-M.~Bonnet-Bidaud$^{6}$, 
M.~Mouchet$^{7}$,
R. Coppejans$^{5,8}$,
A.A.S~Gulbis$^{5}$\\
$^{1}$ INAF $-$ Osservatorio Astronomico di Capodimonte, Salita Moiariello 16, I-80131 Napoli, Italy\\
$^{2}$ Instituto de Astrof\'isica de Canarias (IAC), E-38205 La Laguna, Tenerife, Spain\\
$^{3}$  INAF $-$  Osservatorio Astronomico di Trieste, 	Via G. Tiepolo 11, I-34131 Trieste\\
$^{4}$ Southern African Large Telescope Foundation, PO Box 9, 7935 Observatory, Cape Town, South Africa\\
$^{5}$ South African Astronomical Observatory, PO Box 9, 7935 Observatory, Cape Town, South Africa\\
$^{6}$ CEA Saclay,  DSM/Irfu/Service d'Astrophysique, F-91191, Gif-sur-Yvette, France \\
$^{7}$ Laboratoire APC, Universit\'{e} Denis Diderot, 10 rue Alice Domon et L\'{e}onie Duquet, F-75005 Paris, France \\
$^{8}$ University of Cape Town, Private Bag X3, 7701, Rondebosch, South Africa
}

\begin{document}

\date{Accepted 2014 August 8. Received 2014 August 7; in original form 
2014 March 21}

\pagerange{\pageref{firstpage}--\pageref{lastpage}} \pubyear{2014}

\maketitle

\label{firstpage}

\begin{abstract}
The peculiar low mass X-ray binary XSS\,J12270-4859, 
associated with the \emph{Fermi}/LAT source 2FGL\,J1227.7-4853, was
in a X-ray, gamma-ray and optical low-luminosity persistent state 
for about a decade until the end of 2012, when it has entered into the dimmest state ever
observed.  The nature of the compact object  
has been controversial until the detection of a 1.69ms radio pulsar early 2014.   
We present optical spectroscopy and optical/near-IR photometry 
 during the previous brighter and in the recent faint states.
We determine the first spectroscopic orbital ephemeris  and
an  accurate orbital period of 6.91246(5)\,h. We infer a mid G-type 
 donor star and a distance  d= 1.8-2.0\,kpc.
The donor spectral type changes from G5\,V to F5\,V between inferior and superior conjunction,
 a signature of strong irradiation effects.  We infer 
a binary inclination $45^o \lesssim i \lesssim 65^o$ and  
a highly undermassive donor, $\rm M_2 \sim 0.06-0.12\,M_{\odot}$ for a neutron star mass in the
range 1.4-3\,M$_{\odot}$.  Thus this binary
joins as the seventh member the group of  "redbacks".
 In the high state, the emission lines reveal the presence of an accretion disc. They 
tend to  vanish at the donor star superior conjunction, where also flares are preferentially 
observed together with the occurrence of random dips.  
This behaviour could be related to the propeller mechanism of the neutron star recently 
proposed to be acting in this system during the high state. 
In the low state,  the emission lines are absent at all orbital phases indicating that accretion has completely 
switched-off and that  XSS\,J12270-4859  has transited from an accretion-powered to a rotation-powered phase.
\end{abstract}


\section{Introduction}

Since the discovery of the first radio   millisecond (MSP) pulsar \citep{Backer82}, low-mass X-ray binaries 
(LMXBs) containing neutron stars
(NS) were believed to be possible progenitors of radio  MSPs  through the so-called 
"recycling" scenario \citep{Alpar82}, in which  radio pulsars are spun-up to very short periods during a previous 
prolonged (Gyr) phase of mass accretion. During the mass transfer phase, 
a large amount of angular momentum is transferred to the NS which is spun-up by accretion torques. 
The system appears as a bright accretion powered LMXB.  When mass transfer ceases, the emission is powered by 
the fast rotating NS magnetic field and appears as a radio MSP. Supporting evidences were found from the   
detection of msec X-ray pulses in accreting NS (see for a review \citet{PatrunoWatts}). However 
confirmations of the  link between LMXBs and MSPs came only recently. 
The radio MSP  PSR\,J1023+0038 was found to have  experienced a previous accretion epoch in 2001
\citep{Wang09,Archibald09,Tam,Archibald10} and it is back again in an accretion powered phase since mid 
2013 \citep{Stappers14,Takata14,Patruno14}. Also the MSP, PSR\,J1824-2421, 
in the globular 
cluster M28,  
was surprisingly discovered as an X-ray source during a thermonuclear outburst in early 2013 and, after 
fading in the X-rays, it was again detected as a radio pulsar  \citep{Papitto13}. These discoveries
provide the long-sought "missing-link" and demonstrate that radio MSPs may have episodes of 
mass accretion, supporting the proposals that the transitions between the two states reflect the interplay 
between the NS magnetosphere and the mass transfer rate from the donor star (see 
\citet{Stella94,Campana98,Burderi2001}).
 They furthermore challenge the existence of transitional MSP binaries to be identified yet. In 
particular radio MSPs systems have greatly increased in number with many associated with  \emph{Fermi} sources
\citep{Ransom11,Kaplan12,Kerr12,Kong12,Roberts13} among which PSR\,J1023+0038 \citep{Tam}.

The hard X-ray source XSS\,J12270-4859 (henceforth XSS\,J1227) 
 was early proposed as a cataclysmic variable hosting a magnetic white
dwarf \citep{masetti06,Butters08} and later disproved
with independent observations by \citet{Pretorius09},
\citet{Saitou}, and \citet[][henceforth dM10]{deMartino10}. 
It was identified as a peculiar low-luminosity 
($\rm L_X \sim 6\times 10^{33}\,d_{1\,kpc}^2 erg\,s^{-1}$) 
LMXB with unusual dipping and flaring behaviour (dM10). 
 Surprisingly we also found XSS\,J1227 to be positionally 
coincident with the high energy gamma-ray \emph{Fermi}-LAT source 
1FGL\,J1227.9-4852/2FGL\,J1227.7-4853 emitting up to 10\,GeV (dM10).
Multi-wavelength follow-up observations, performed by \citet{Hill}, 
\citet[]{Saitou11} and \citet[][henceforth dM13]{deMartino13},  
showed  persistent X-ray and high energy gamma-ray emissions
over an interval of at least $\sim$ 7\,yr, 
making it an unique case among known galactic binaries with comparable
X-ray and gamma-ray luminosities. 
The detection of a radio \citep{Hill} and gamma-ray (dM10) counterpart, 
together with a putative 4.5\,h orbital period (dM10), led to the suggestion that 
XSS\,J1227 could host a MSP \citep{Hill}, 
sharing similar properties with PSR\,J1023+0038 \citep{Wang09,Tam,Archibald10}. However, no radio 
pulses 
were detected \citep{Hill} and searches for msec X-ray pulses only provided  
upper limits to pulse amplitudes of $\sim 15-25\%$  (dM13). The
nature of the compact object, an unusual black hole emitting in high energy 
gamma-rays or a peculiar state of a MSP, remained an open question. The 
peculiar persistent optical, X-ray and high energy gamma-ray emissions were 
recently discussed by \citet[][]{Papitto14} in terms of a fast spinning  
NS in a propeller state. The propeller effect operates 
when the magnetospheric radius is larger than the corotation radius thus inhibiting 
 accretion of matter. The  model predicts that, at low accretion rates, the accretion
flow is truncated at large distances  by the magnetic field of a rapidly 
rotating NS, where electron acceleration occurs producing 
energetic gamma-rays. This model also allowes a small fraction of the inflowing matter to
be accreted onto the NS. 

 \noindent XSS\,J1227 remained stable in gamma-rays, X-rays and optical 
for about a decade until the end of 2012, when it 
decreased by $\sim$ 1.5-2\,mag in the optical and by more than a factor of 
ten in the X-ray flux  \citep{Bassa13,Bassa14}.    Hints of a decrease in the \emph{Fermi}-LAT 
gamma-ray flux, coincident with the optical decay were 
also found by \citet{Tam13}.  This behaviour prompted
 the suggestion that XSS\,J1227 could be 
transiting from a LMXB to a MSP powered phase \citep{Bassa13} 
reminiscent of the state transitions seen in the binary MSPs 
PSR\,J1824-2421  and PSR\,J1023+0038 itself. The unexpected 
variability triggered further follow-ups, among which radio searches,  
at 607\,MHz with GMRT early 2014, finally detecting 
a highly accelerated 1.69\,ms pulsar \citep{Roy14}.  

\noindent XSS\,J1227 has still to be characterised
in terms of system pararameters and energetics to understand its peculiar properties.
The early claim of a  4.5\,h photometric periodicity (dM10) 
was subsequently discarded by 
U-band \emph{XMM-Newton}/OM observations giving a longer period 
(6.4(2)\,h) (dM13). A 6.91\,h periodicity was found from a preliminary 
analysis of part of the spectroscopic data presented in this work 
\citep{deMartino13b,deMartino13c} and further confirmed  by \emph{Swift}/UVOT U-band
data \citep{Bassa14}.

We present here the results of optical spectroscopic and 
optical/near-IR (nIR) photometric campaigns
carried out in early 2012 during the long persistent high state 
and in late 2013  during the faint state into which the source entered 
since the end of 2012.

\noindent In Sect.\,2 we describe the spectroscopic and photometric 
observations and data reductions. In Sect. 3 we present the spectroscopic 
analysis from which we detect both donor and compact star orbital motions 
and derive  the first spectroscopic orbital period and ephemeris. In 
Sect.\,4,  we present the  peculiar photometric variability and link it 
 to a complex spectral behaviour during the high state as well as
the orbital variability during the dim state. In Sect.\,5 we discuss
the results, obtaining estimates of the binary parameters, the distance 
and the effects of irradiation on the secondary star. We finally discuss
the accretion flow and peculiar variability in terms of a propeller state.

\section[]{Observations and data reduction}

\subsection{The spectroscopic data}

We observed XSS\,J1227 from 2012, Mar. 30 to Apr. 1, with the 4m NTT
telescope at the ESO La Silla Observatory equipped with the EFOSC2 spectrograph 
using grism\#19 (1557\,l/mm) that 
covers the spectral range 4445-5110\,$\rm \AA$ with  spectral resolving power
 R$\sim$ 2200--3000.
A total of 73 spectra with 
exposure time of 1200\,s were obtained. During each night we took arc-lamp exposures every 
$\sim$1-1.5\,h
to compensate for possible shifts due to instrumental flexure. 
Spectrophotometric standard stars were also observed at the beginning 
(LTT\,3864 and GD\,108) or at  the end (EG\,274) of  each night.

\noindent Extended spectroscopic coverage was secured  
 from 2012, Mar. 28 to Apr. 2 at the 10\,m Southern 
African Large Telescope (SALT) \citep{Buckley06}
 equipped with the Robert Stobie Spectrograph (RSS)\citep{Burgh03}  
and grating\,PG2300 (2300\,l/mm) 
covering the range 4000-5000\,$\rm \AA$  with  R$\sim$2200-5000. 
A total of 65 spectra  were obtained with integration time of 300\,s. 
 
\noindent Further spectra were acquired from 2013, Dec. 14 to Dec. 18 
at the ESO/NTT with the same instrument set-up. A total of 7 spectra 
with exposure times
between 900\,s and 1200\,s were obtained. When two spectra within 
the same night were acquired, they are separated by more than 2\,h in order 
to sample different orbital phases. 
The log of the spectroscopic observations is reported in Table\,\ref{observspec}.

\noindent The NTT data reduction was performed following standard 
IRAF procedures,  using bias and flat field frames.  Wavelength calibration of the 2012 data set was 
performed using the arc exposures so that every three
NTT spectra had its own wavelength calibration providing 
 a spectral resolution (FWHM) of 1.55\,$\rm \AA$ (96\,km\,s$^{-1}$) and 4.3$\rm \AA$ (265\,km\,s$^{-1}$)
for the 0.7" and 1.0" set-ups respectively. On the other hand, only one
arc-line spectrum was acquired on 2013 Dec. 14 giving a 
spectral resolution of  4.4\,$\rm 
\AA$ (270\,km\,s$^{-1}$). 
Flux calibration for all NTT spectra was performed 
using the sensitivity function created by combining all the standard stars 
and correcting for gray losses when necessary.

\noindent The SALT spectra were processed with the PySALT pipeline \citep{Crawford2010}
to perform  gain and cross-talk corrections, overscan bias subtraction and mosaicing. 
Standard IRAF procedures were employed to perform flat field correction, 
wavelength  calibration and background subtraction.  
%Cosmic rays, routinely cleaned by the pipeline
%using light parameters,  and as a consequence a few left were removed interactively.
The wavelength calibration was based on arc exposures acquired at the 
beginning and end of each SALT observing slot providing a resolution of  
$\sim$2.7\,$\rm \AA$ (165\,km\,s$^{-1}$).  We further applied a 
shift\footnote{A systematic  velocity  offset of unknown origin between the NTT and SALT 
$\gamma$ velocities was found.  The shift, currently under investigation, is however smaller than the 
SALT/RSS resolution element.} 
of 33.9\,km\,s$^{-1}$ to match those obtained with NTT. No 
spectrophotometric  standard star was observed during the SALT run 
and therefore the spectra were not flux calibrated.

\noindent Heliocentric time and velocity corrections were applied to both 
sets of data.

\subsection{The photometric data}

XSS\,J1227 was observed from Mar. 28 to Apr. 1 with the 1.9m telescope
at the South African Astronomical Observatory (SAAO) equipped with the 
Southerland High-speed Optical Camera (SHOC) \citep{Coppejans}  
during the commissioning phase, using the U band filter and adopting a 5s integration  
time. The U filter  was choosen for comparison of with 
the 2009 and 2011 light curves obtained with the  \emph{XMM-Newton} Optical Monitor 
(dM10,dM13). The photometric coverage is $\sim$ 4.4\,h,
7.7\,h and 7.4\,h during the three nights, respectively. 

\noindent Simultaneous nIR photometry in J (1.25$\mu$), 
H (1.63$\mu$) and K$\rm_s$ (2.14$\mu$) bands was also acquired 
from Mar. 28 to Apr. 3  at the SAAO 1.4m InfraRed Survey Facility (IRSF) 
telescope  equipped with the 
Simultaneous three-color InfraRed Imager for Unbiased Survey 
(SIRIUS) \citep{Nagayama03}. 
15\,s exposures  with a dithering at 10 positions were performed giving a temporal resolution
of 256\,s including deadtimes.  During the five nights the source was observed for
7.0\,h, 9.8\,h, 8.1\,h, 10.7\, and 9.6\,h respectively. 
The optical and nIR photometry  
partially  overlaps the NTT and SALT spectroscopic observations. 

\noindent  XSS\,J1227 was re-observed from 2013 Dec. 13 to Dec. 15 with the 
0.5m REM telescope equipped with the ROSS camera simultaneously in  g' and r' bands, 
using integration times of 120s.  
Total coverages of 1.9\,h, 0.7\,h and 
0.8\,h were obtained during the three nights.  It was also simultaneously 
observed  with the REMIR camera in the J filter but its faintness 
precluded  useful analysis.
The log of the photometric observations is reported in Table\,\ref{observphot}.

\noindent  All photometric data sets were reduced using standard routines of 
IRAF to perform 
 bias and flat-field  corrections. For the IRSF/SIRIUS photometry fringe 
patterns by OH were eliminated and the 
ten dithered images were merged into a single frame.
For each data set, aperture photometry was performed optimizing 
aperture radius and sky subtraction was done using annuli of different 
sizes.  
Comparison stars  were used to check  and to correct for variable 
sky conditions.  No photometric calibration was applied to the 1.9m/SHOC 
data and therefore we present differential photometry.
 The IRSF/SIRIUS JHK magnitudes 
were reconstructed using the comparison 2MASS star at: 
RA$_{2000}$ =12:28:01.885, DEC$_{2000}$= -48:52:09.77, with 
J=13.076 $\pm$ 0.033 H=12.710 $\pm$ 0.037 and K$_{\rm s}$ =12.670$\pm$ 0.031. 
The REM/ROSS photometry was calibrated 
 using the Sloan standard SA\,92\,345 observed during the three nights.  
Heliocentric corrections were applied  to all time series.

\begin{table*}
 %\centering
\flushleft
 \begin{minipage}{140mm}
  \caption{\label{observspec} Summary of the spectroscopic observations}
  \begin{tabular}{@{}llllccccc@{}}
  \hline
Telescope/Instrument &   UT Date  & UT Start & \multicolumn{4}{c}{Instrument setup} & exptime & \# exposures \\
%& sky & seeing$^{\dagger\dagger}$ ($^{\prime\prime}$) \\
          &  &  &  grism & slit & CCD binning & dispersion               & (sec) &    \\
          &  &  &        & (")  &             & ($\rm \AA$/pix)          &  & \\
 \hline
%NTT/EFOSC2 & 2012-03-30 & 00:04 & \#19 & 0.7 & 2$\times$1$^\dagger$ & 0.33 & 1200 & 21 & CLR & 0.6-1.2\\
ESO NTT/EFOSC2 & 2012-03-30 & 00:04 & \#19 & 0.7 & 2$\times$1$^*$ & 0.33 & 1200 & 21 \\
           & 2012-03-30 & 08:17 & \#19 & 1.0 & 2$\times$2 & 0.67 & 1200 & 4 \\
           & 2012-03-31 & 00:37 & \#19 & 1.0 & 2$\times$2 & 0.67 & 1200 & 23 \\
           & 2012-04-01 & 00:04 & \#19 & 1.0 & 2$\times$2 & 0.67 & 1200 & 25 \\
%           & 2012-04-01 & gr\#5  & 1.0 & 2$\times$2 & 4.04 & 600 & 1 & CLR & 1.5\\
%& & & & & & & & & \\
           & 2013-12-14 & 07:21 & \#19 & 1.0 & 2$\times$2 & 0.67 & 900 & 2 \\
           & 2013-12-15 & 07:35 & \#19 & 1.0 & 2$\times$2 & 0.67 & 900 & 1 \\
           & 2013-12-17 & 07:12 & \#19 & 1.0 & 2$\times$2 & 0.67 & 1200 & 2 \\
           & 2013-12-18 & 06:12 & \#19 & 1.0 & 2$\times$2 & 0.67 & 1200 & 2 \\
&  & & & & & & \\
%\hline
%Robert Stobie Spectrograph 
SAAO SALT/RSS   & 2012-03-28 & 19:12 & PG2300 & 1.5 & 2$\times$2      & 0.35 & 300 & 13\\ 
           & 2012-04-01 & 01:04 & PG2300 & 2.0 & 4$\times$8           & 0.68 & 300 & 11\\
           & 2012-04-01 & 18:52 & PG2300 & 2.0 & 4$\times$8           & 0.68 & 300 & 14\\
           & 2012-04-02 & 00:56 & PG2300 & 2.0 & 4$\times$8           & 0.68 & 300 & 13\\
           & 2012-04-02 & 18:51 & PG2300 & 2.0 & 4$\times$8           & 0.68 & 300 & 14\\
 %     &  &  &  &  &  &  &  &  & \\
\hline
\end{tabular}
$^*$ the binning is in the spatial direction.\\

\caption{\label{observphot} Summary of the photometric observations}
\begin{tabular}{@{}lllccc@{}}
  \hline
Telescope/Instrument &   UT Date  & UT Start & Filter & exptime & \# exposures \\
& & & (sec) & & \\
\hline
SAAO 1.9m/SHOC & 2012-03-28 & 21:49 & U      & 5      & 3159\\
             & 2012-03-31 & 19:05 & U      & 5      & 5573\\
             & 2012-04-01 & 18:39 & U      & 5      & 5320\\
& & & & & \\
SAAO IRSF/SIRIUS & 2012-03-28 & 20:09 & J,H,K & 150$^{**}$ & 99\\
                 & 2012-03-31 & 18:09 & J,H,K & 150 & 138\\
                 & 2012-04-01 & 19:46 & J,H,K & 150 & 124\\
                 & 2012-04-02 & 17:27 & J,H,K & 150 & 156\\
                 & 2012-04-03 & 17:39 & J,H,K & 150 & 150\\
& & & & & \\
ESO REM/ROSS         & 2013-12-13 & 07:04 &   g',r'  &  120 & 24\\ 
                 & 2013-12-14 & 07:03 &  g',r'   &  120 & 12\\
                 & 2013-12-15 & 06:11 &   g',r'   &  120 & 14\\
& & & & & \\
\hline
\end{tabular}

$^{**}$ 15s exposures in dithering mode at 10 positions\\ 

\end{minipage}
\end{table*}

\begin{figure}
\includegraphics[width=3.0in, clip=true, trim=0 5 0 5]{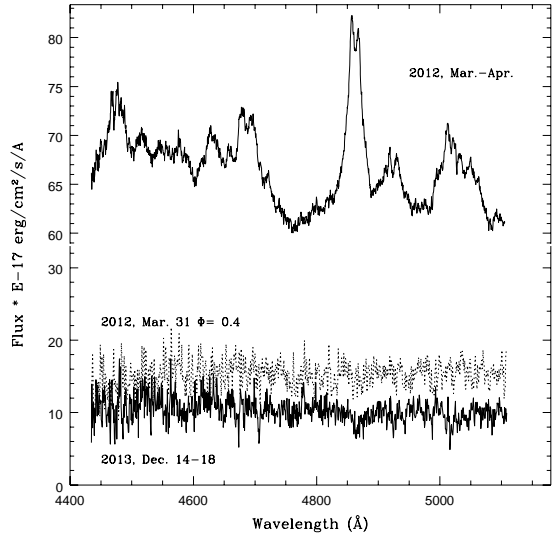}
 \caption{The grand average spectra of XSS\,J1227 obtained at ESO/NTT in 2012 (top) 
and in 2013 (bottom)  showing the dramatic changes between the two epochs. 
Such spectacular changes are also observed in 2012 on timescales of hours 
as delucidated by the spectrum (dotted line) acquired at orbital 
phase $\sim$0.4 (see text).}
\label{xss_ave_all}
\end{figure}

\section{Results from spectroscopy}

The spectra obtained in 2012 show XSS\,J1227 at an average flux level 
similar to 
that observed in 2006 by \citet{masetti06},  with prominent 
H$_{\beta}$, He\,I ($\lambda\lambda$ 4471, 
4919, 5015), He\,II ($\lambda\lambda$ 4686) and bowen-blend NIII 
($\lambda\lambda$ 4634-41)/CIII 
($\lambda\lambda$  4647-50) (Fig.\,\ref{xss_ave_all}, top).  A strong 
variability is found in the continuum and emission lines.
Dramatic changes from an emission dominated spectrum to a pure absorption 
one on  timescales of hours were recorded in both NTT and SALT spectra.
These spectacular changes are also accompained by strong flux variations 
(Fig.\,\ref{xss_ave_all}, middle).  
Several absorption features of metallic lines are detected in the individual
exposures even when emission are present. These are undoubtedly ascribed 
to the donor star. The relative strength of metallic lines suggests the companion is
approximately an F-type star. Wavelength shifts in emission and absorption lines are 
signature of orbital variability.

\noindent In comparison, the spectra obtained in 2013 do not 
show any emission feature  
(Fig.\,\ref{xss_ave_all}, bottom) but rather weak absorptions
overimposed on a faint continuum, a factor of $\sim$7 lower than 
that observed in 2012 and 2006.
The average flux is 
$\rm \sim 1.1\times 10^{-16}\,erg\,cm^{-2}\,s^{-1}\,\AA^{-1}$ comparable to
only one spectrum acquired on 2012, Mar. 31 when emission lines were 
also absent (middle panel). The source appears to have undergone into a deep low state with  
no sign of accretion.

\subsection{The orbital variability}

\begin{figure}
\begin{tabular}{c}
\includegraphics[width=2.0in, angle=-90]{xss_fig2a.ps}\\
\includegraphics[width=3.0in, angle=0]{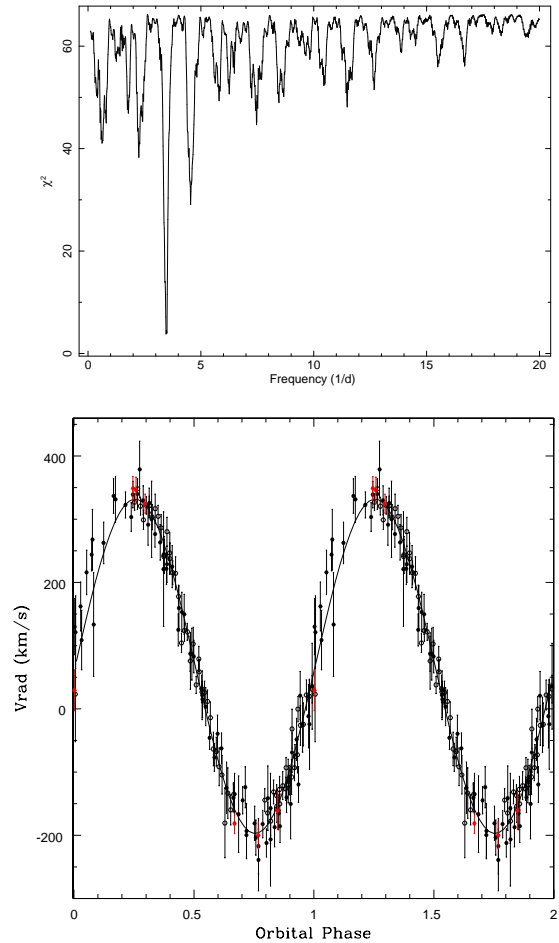}\\
\end{tabular}
 \caption{{\it Top:} The periodogram of radial velocities obtained 
by cross-correlating the NTT and SALT 
spectra acquired in 2012 with a F2\,V template broadened  by 
125\,km\,s$^{-1}$. {\it Bottom:} The radial velocities  folded at  the refined 
orbital  period using the ephemeris quoted in the text together with the 
best fit 
sinusoid.  Measures from NTT are marked with filled circles (black from 
2012 run and red from 2013 run), 
from SALT with  empty circles.}  
\label{period_fold}
\end{figure}

We cross-correlated the individual NTT and SALT spectra acquired in 2012 with
the F2 V template HR 2085\footnote{The template spectrum was obtained with the ISIS  
spectrograph on the 4.2m William Herschel Telescope (WHT) on the night of 2 Dec 1997 at 
64 km $s^{-1}$ resolution \citep{Casares98}.} 
 in the spectral ranges $\lambda\lambda$4500-4615, 
4725-4820, 4940-4990 and 5065-5010, free from strong emission lines. 
The template was broadened to
125 km $s^{-1}$ to match the width of the absorption features in XSS\,J1227 
(see below). The radial velocities were obtained by parabolic fit to the peaks
of the cross-correlation functions. The systematic offset  between the SALT and NTT 
velocities was corrected by shifting the SALT spectra by +33.9 km $s^{-1}$. 
A power spectrum analysis of the radial velocities reveals a clear peak at a 
frequency of  3.47 cycles d$^{-1}$, corresponding to 0.288\,d  (Fig.\,\ref{period_fold}, 
top panel). A 
sine wave fit to the radial velocities yields P$_{\rm orb}=0.28775(17)$d, the time of 
blue-to-red zero crossing T$_o=2456015.1574(20)$ (inferior conjunction of the companion 
star), the  amplitude of the donor star K$_2=261(5)$ km s$^{-1}$ and the systemic velocity
$\gamma=67(2)$ km s$^{-1}$, where errors are purely statistical.  We have 
rescaled the errorbars of the NTT and SALT
measures by factors of 2.5 and 2.7, respectively, in order to obtain a reduced
$\chi^2=1$. Three outliers were masked from the fit, although they have a
negligible impact in the final parameters.

\noindent We also extracted radial velocities from the 2013 NTT spectra using the same
template. These velocities are clearly modulated with the 0.288\,d orbital period.
An offset of -87 km s$^{-1}$ is detected,  most likely caused by flexure
differences between the target observations and the calibration arc 
obtained during daytime with the telescope pointing to the zenith. Therefore, 
the 2013 NTT velocities were corrected by adding 87 km s$^{-1}$. A sine wave  
fit to these velocities, fixing  P$_{\rm orb}=0.28775$d, yields a more
contemporary determination of the zero phase, T$_o=2456640.736(5)$. The  
uncertainty in P$_{\rm orb}$ prevents us from constraining the exact number of
orbital cycles elapsed betwen the two zero phases and, therefore, improve 
the orbital period. A periodogram of the entire database of radial
velocities is dominated by a strong aliasing pattern with several peaks
around 3.475 cycles d$^{-1}$ equally significant. However, only one peak
(at 3.472 cycles d$^{-1}$) is consistent with the photometric period reported 
by \citep{Bassa14}.  Assuming the photometric period is the orbital period and not the superhump (e.g. 
\citep{Haswell01}, a sine wave fit to the radial velocities, 
using Bassa et al's period as input parameter, allowes us to refine the period to     
P$_{\rm orb}=0.2880195(22)$\,d. Our best set of orbital parameters are reported
in Table\,\ref{orbparam}, while the radial velocity curve, folded on this 
period, is presented in the bottom panel of  Fig.\,\ref{period_fold}.

\begin{table}
%\centering
\flushleft
% \begin{minipage}{140mm}
  \caption{\label{orbparam} Spectroscopic orbital parameters of XSS\,J1227}
  \begin{tabular}{lc}
  \hline
& \\
P$_{\rm orb}$ & 0.2880195(22)\,d \\
K$_2$ & 261(5)\,km\,s$^{-1}$ \\ 
K$_1$ &  89(23)\,km\,$^s{-1}$ \\
$\gamma$ & 67(2)\,km\,s$^{-1}$ \\
T$_o^{\star}$  & 2456015.1574(20) \\
T$_o^{\star\star}$  & 2456640.736(5)\\
& \\
  \hline

\end{tabular}

$^{\star}$ Time of blue-to-red zero crossing of 2012 data set \\
$^{\star\star}$ Time of blue-to-red zero crossing of 2013 data set \\

%\end{minipage}
\end{table}

\begin{figure}
\begin{tabular}{c}
\includegraphics[width=2.0in, angle=-90]{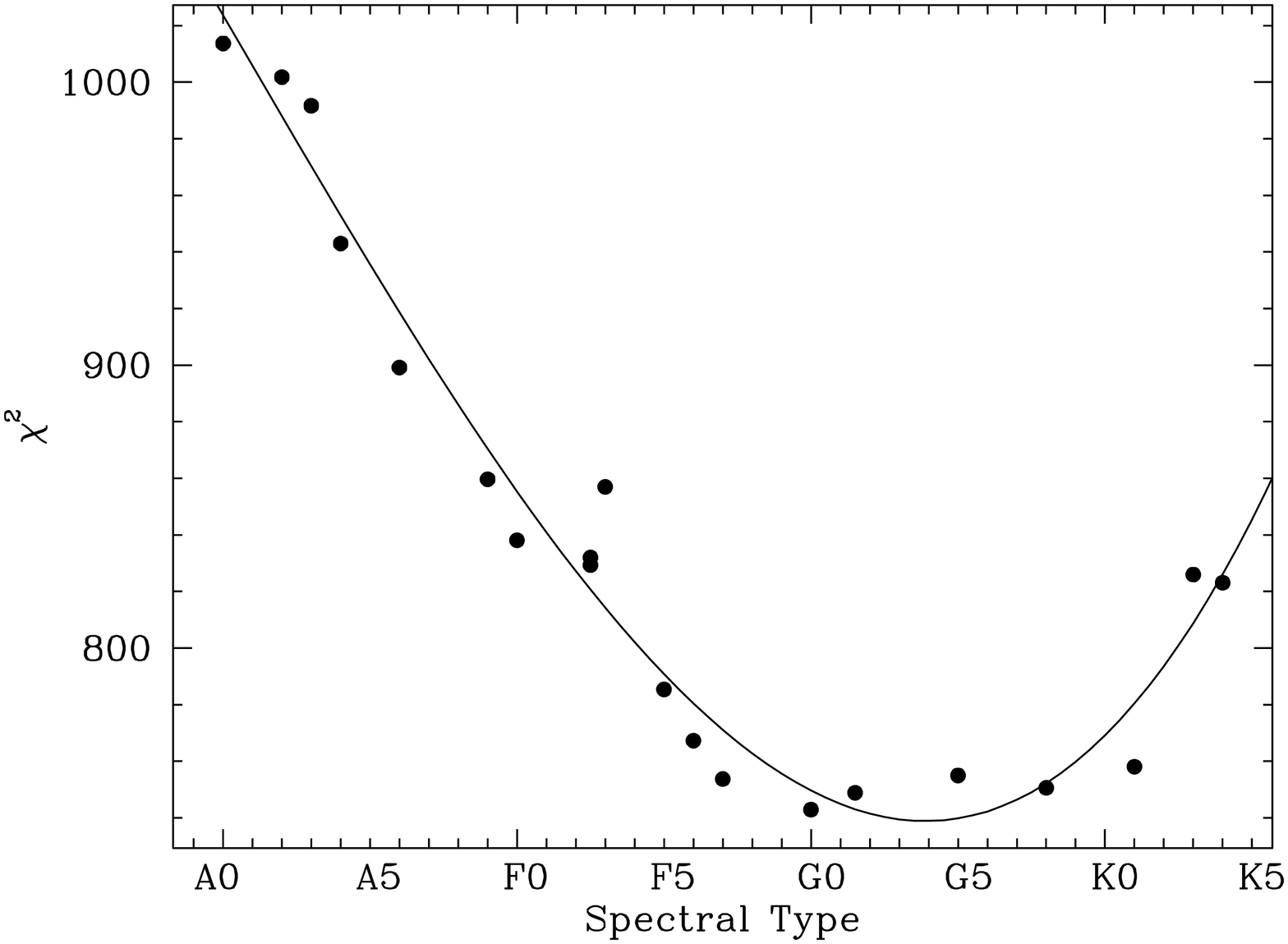}\\
\includegraphics[width=2.0in, angle=-90]{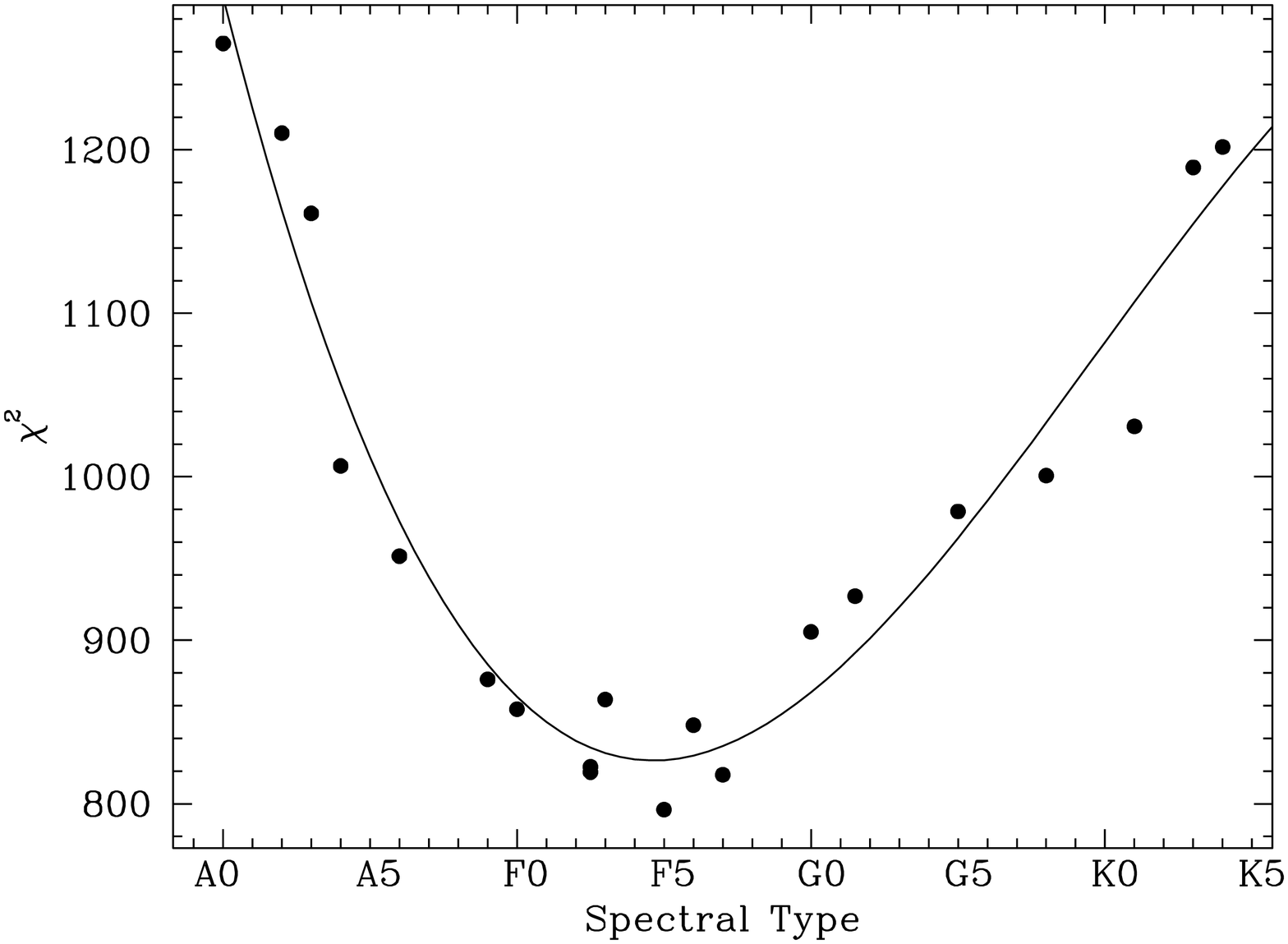}\\
\includegraphics[width=2.0in, angle=-90]{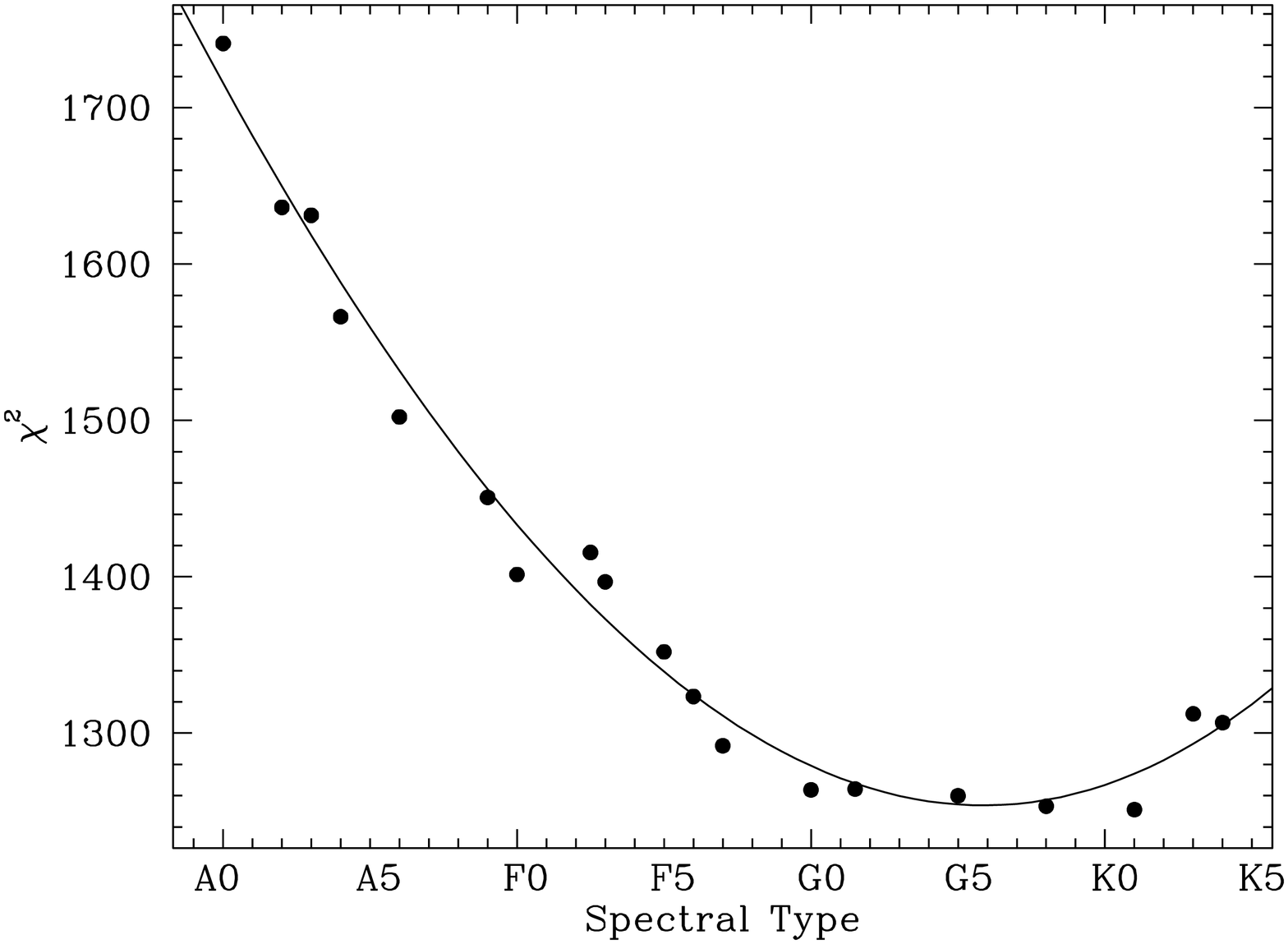}\\
\end{tabular}
 \caption{The $\chi^2$ distribution from cross-correlation of average spectra obtained in 2012
around orbital phases 0.0 (top) and 0.5 (middle) and in 2013 (bottom). 
Abscissas indicate spectral types  A0 (0), F0 (10), G0 (20) and K0 (30). 
The spectra in 2012 show earlier spectral type at the secondary superior 
conjunction. A mid G-type star is favoured from 2013 data.}
\label{chi2sptype}
\end{figure}

\begin{figure}
\begin{tabular}{c}
\includegraphics[width=2.0in, angle=-90]{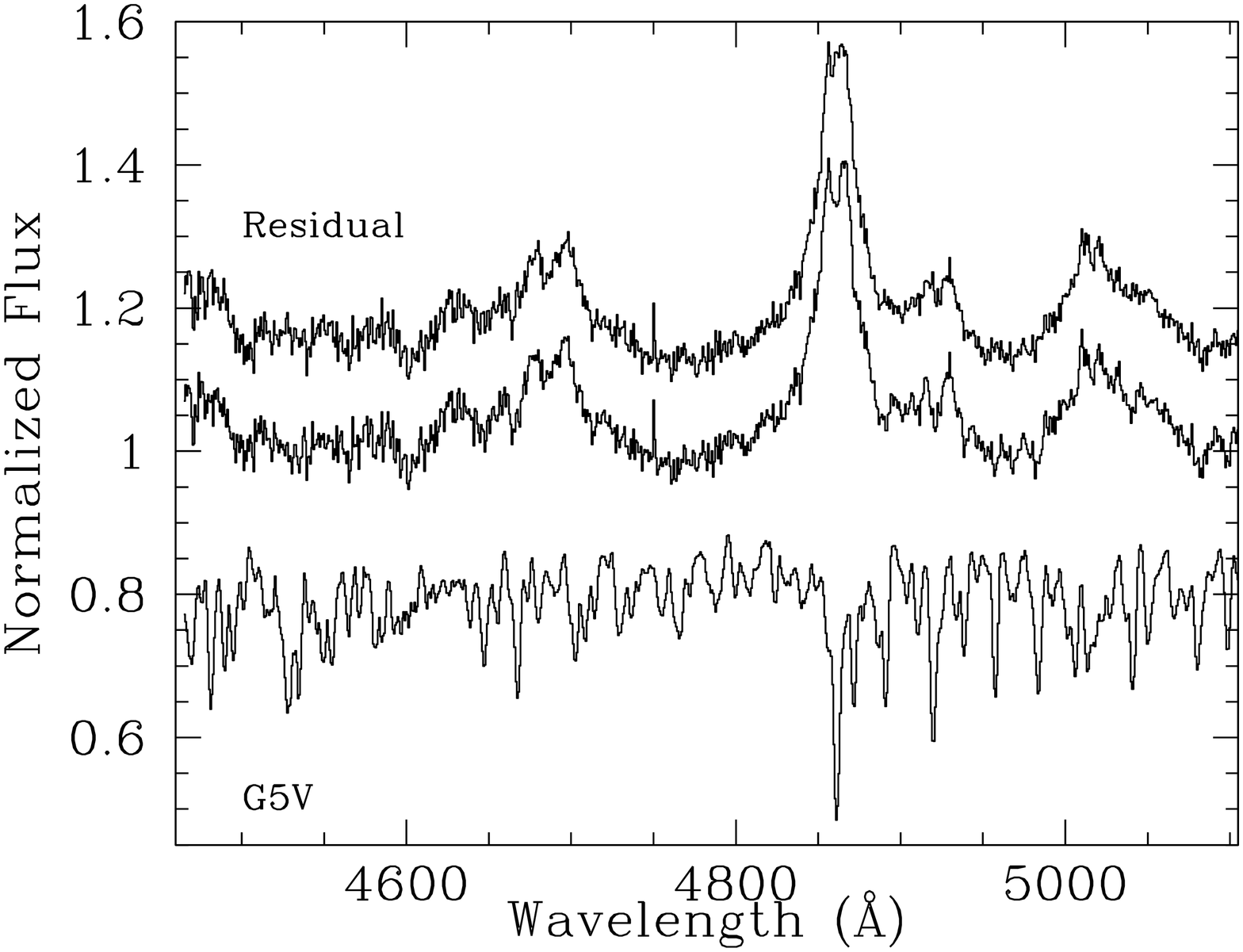}\\
\includegraphics[width=2.0in, angle=-90]{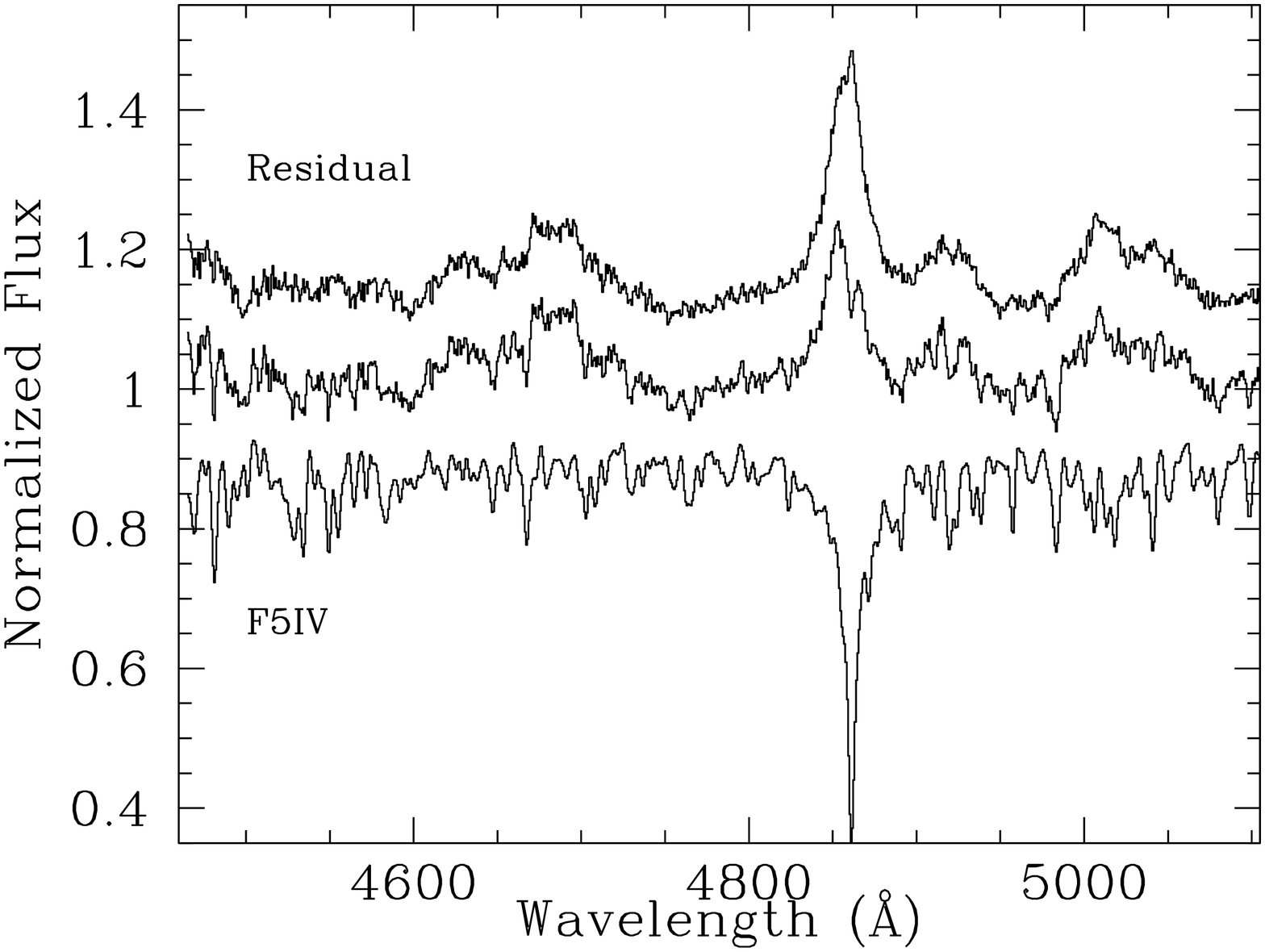}\\
\end{tabular}
 \caption{ The Doppler corrected average NTT spectra around orbital phase 0.0 (top) and 0.5 (bottom)
together with the corresponding best fit templates, G5\,V and F5\,V, respectively. The residual 
spectra after template subtraction are also shown (shifted vertically for clearness).}
\label{sptype}
\end{figure}

\noindent 
 Variations in the depth of several metallic lines with orbital phase
were also detected,  indicating  irradiation effects. To test this, we 
produced
Doppler corrected averages in the reference frame of the companion star
(using the ephemeris in Table\,\ref{orbparam}) grouping spectra at phases 0.9-1.1 and 
0.4-0.6. These were compared with a set of A0-F7 templates of luminosity class IV-V from
\citet{Casares98}, conveniently broadened to match the XSS J1227 spectra.
These were complemented with another set of high-resolution main sequence
templates with spectral types G0-K7 (see details \citep{Casares09}) 
The spectral type classification of the XSS\,J1227 spectra was
obtained through an optimal subtraction routine (see e.g. \citet{Casares09}) 
and the results are presented in the top and middle panels of Fig.\,\ref{chi2sptype}. 
At  orbital phase 0, the minimum $\chi^2$  is found at
 spectral type G5 whereas it is shifted to F5 at phase 0.5. The Doppler corrected averages at
the two phases are shown in  Fig.\,\ref{sptype} together with their respective best fit templates 
and the residuals after template subtraction. Here we note that 
the spectrum at phase 0.5, averaged over the three nights,  shows emission lines due to
the night-to-night variability (see below and Sect.\,3.2). 
The orbital dependence of spectral type is a clear indication that the secondary star is affected 
by irradiation at superior conjunction.  This may cause some quenching on the photospheric 
absorption lines \citep{Martin89} which could lead to an  overestimate of 
the K$_2$ amplitude (see Sect. 5.1). We also attempted to constrain the spectral type of 
the companion star from the 2013 NTT spectra. Here, because of the scarce number of 
spectra and their lower signal-to-noise, we produced a single Doppler corrected average 
combining the 6 spectra from all phases. The best match is obtained for the G6\,V
template, although  spectral types between G0 and K1 are equally
significant (Fig.\,\ref{chi2sptype}, lower panel). The data confirm that the donor is a
mid-G star, which is heated up to spectral type $\sim$ F5.

\noindent To estimate the companion star rotational 
broadening we  used a subset of 21 NTT spectra from the 
first night in 2012, that  have better spectral resolution. 
These  were coadded in the rest frame of the companion star and the resulting spectrum 
was compared with an F7\,IV and a G0\,V template, both  degraded to the  
resolution of the NTT spectra (96 km\,s$^{-1}$) and subsequently broadened from 50 to 
100 km\,s$^{-1}$ in steps of 1 km\,s$^{-1}$ (see e.g. \citet{Casares10}).
The best match is obtained when the F7\,IV is broadened
broadened by   98$\pm$8\,km\,s$^{-1}$ and the G0\,V template to 
82$\pm$16\,km\,s$^{-1}$ give equally good fits.  Therefore, we  conservatively 
adopt  $V\,sin\,i$= 86$\pm$20\,km\,s$^{-1}$, although this number should be treated
with caution because it is comparable to our instrumental resolution.

\begin{figure} 
\includegraphics[width=3.0in, angle=0]{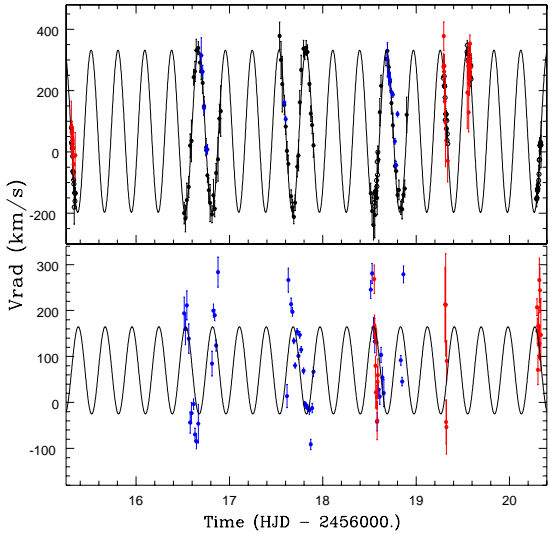}
\caption{{\it Top:} The radial velocities 
as obtained from cross-correlation of metallic lines
with a F2\,V template (NTT:black filled and SALT open black circles)  and from gaussian fits to the
H$_{\beta}$ absorptions as measured on the NTT (blue circles) and SALT spectra (red circles).
The sinusoidal fit using the orbital solution quoted in the text is also shown.
  {\it Bottom:} The
radial velocity measures as obtained from Gaussian fits to the H$_{\beta}$ 
emission line on the NTT
(blue circles) and SALT (red circles) together with a sinusoidal fit. Note the
night-to-night variability of emission lines.
The radial velocities of emissions and absorptions are anti-phased.} 
\label{vel_time} 
\end{figure}

\noindent To study the emission line behaviour observed in 2012 we 
restricted the analysis of radial velocities  to the H$_{\beta}$ line 
that is the strongest line in the spectrum. The 
profiles were normalised to their local continuum 
and then fitted with  a broad (FWHM$_{em}$ $\sim$ 30\,$\rm \AA$ = 
1850\,km\,$^{-1}$) Gaussian  in emission. 
When in absorption, H$_{\beta}$ profiles were fitted with a narrow 
Gaussian (FWHM$_{abs} \sim$ 3 \,$\rm 
\AA$ = 185\,km\,$^{-1}$). The latter is consistent with the derived rotational 
broadening. A period search was performed on the radial velocity 
measures of H$_{\beta}$ emissions 
from both NTT and SALT data sets, giving a  period of 0.38(1)\,d, 
the 1-d alias of that obtained from cross-correlation of absorption lines. 
This is likely due to the  significant night-to-night variability of radial velocities 
of the emission line with shifts up to 300\,km\,s$^{-1}$ (see Fig\,\ref{vel_time}).   
We therefore fixed the period at the refined orbital value and fit the 
radial velocities  of H$_{\beta}$  emissions and absorptions  obtaining 
K$_{abs}$ = 264(3)\,km\,s$^{-1}$,
$\gamma_{abs}$ = 67(2)\,km\,s$^{-1}$ and K$_{em}$ = 89(23)\,km\,s$^{-1}$, 
$\gamma_{em}$ = 63(10)\,km\,s$^{-1}$. 
The $\gamma$ velocities of absorptions and emissions well
agree within their uncertainties. 
Also, the H$_{\beta}$ absorption radial velocity 
amplitude  matches that derived from cross-correlation.  
The radial velocity curves versus time, together with their corresponding  sine wave 
fits, are shown in Fig.\,\ref{vel_time}. This figure elucidates 
the  two-fold behaviour of H$_{\beta}$ line: the tendency of the emissions
to disappear around orbital phases $\sim$ 0.35--0.65 with a strong 
night-to-night variability, which reveal  the donor photospheric absorptions, 
and the anti-phased modulation of emissions and
 absorptions, which then trace the primary and donor star orbital motions
respectively.

\noindent We further inspected the variability in both continuum and 
H$_{\beta}$ emission fluxes in the NTT spectra, measuring 
the  line integrated and the local continuum fluxes. Both are variable, 
being H$_{\beta}$ flux strongly dependent on the orbital period. Equivalent
widths (EWs) vary by $\sim 90\%$ and are maxima at inferior conjunction of the
donor star.

\subsection{Trailed and Doppler maps}

\begin{figure*}
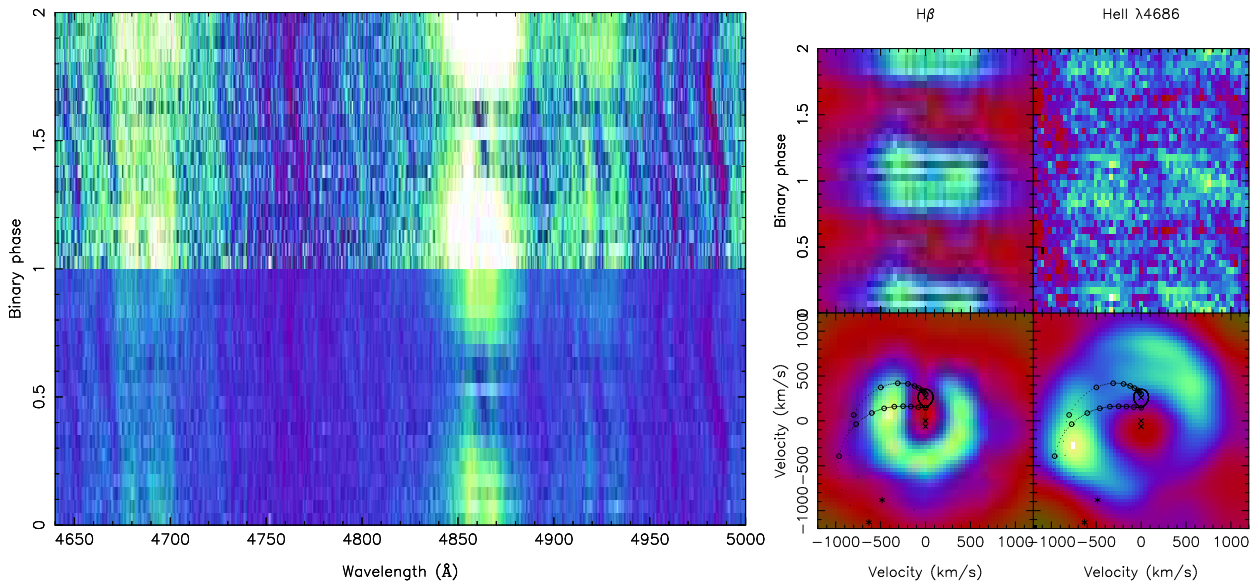

%\begin{tabular}{c}
\includegraphics[width=3.0in, angle=-90]{xss_fig6a.ps}
\includegraphics[width=3.0in, angle=-90]{xss_fig6b.ps}
%\end{tabular}
 \caption{{\it Left:}  The NTT trailed spectra averaged in 20 orbital phase bins revealing the
orbital evolution of several absorption features and emission lines. Two 
color scales have been used  to enhance the absorptions and emission line 
structure. {\it Right:} Enlargments of the regions at H$_{\beta}$ and He\,II  
showing the double-peaked shape (top). Doppler maps in velocity 
coordinates are shown together with the 
theoretical keplerian trajectory along the  gas stream and the stream 
trajectory. The secondary Roche lobe is also  indicated for $q=0.25$}
\label{trail_ntt}
\end{figure*}

Trailed spectrograms were constructed to investigate in more detail the
orbital line variability. We used the NTT set  
which has a more homogeneous coverage and averaging  the spectra 
in 20-phase bins. In Fig.\,\ref{trail_ntt} (left panel) the trailed spectra 
in the region  encompassing He\,II and H$_{\beta}$ are  displayed with 
different color scales to enhance either the metallic 
absorption lines or the peak of the emission lines. The absorptions are clearly 
visible  throughout the orbital cycle, while emission lines 
fade in the phase range $\sim$ 0.35--0.65. This fading is observed in the
spectra during the three nights where also a night-to-night variability in 
the emissivity is detected. 
In Fig.\,\ref{trail_ntt} (right panel) the H$_{\beta}$ trailed profiles (top) reveal a 
double-peaked  structure, a signature of an accretion disc.
 It is worth noticing that the red side of H$_{\beta}$  profile  tends to  disappear
first (left panel of Fig.\,\ref{trail_ntt}), likely due to the increasing contribution of photospheric
absorptions of the donor star.

\noindent Doppler maps of H$_{\beta}$  and He\,II, the two strongest lines in our 
wavelength range, were produced from the 73 NTT spectra following  \citet{Marsh_Horne88}. 
These spectra were continuum subtracted and rebinned into a velocity scale of 
42\,km\,s$^{-1}$ per pixel. The maps are displayed in the right and lower panels of 
Fig.\,\ref{trail_ntt}.  They also display the Roche lobe of the donor star,
together with the ballistic trajectoty of the gas stream and the keplerian
velocity of the disc, in steps of 0.1\,R$_{L1}$ for $q$=$\rm M_2/M_1=0.25$. The presence of   
an accretion disc is further confirmed by the maps. The lack of emission at particular 
azimuths is likely due to the visibility of the strong photospheric H$_{\beta}$ 
absorption from the donor.  Difference between
the two maps are found. The He\,II  displays emission at
larger velocities than H$_{\beta}$,  indicating that this line is formed
closer to the compact star. The   H$_{\beta}$  map shows a bright spot consistent
with emission from the impact region of the gas stream with the outer rim of the disc
at  $\sim$0.55 R$_{L1}$. The latter sets constraints on the outer disc radius.
Instead, the He\,II map shows a bright spot along the gas stream trajectory but further in, at 
$\sim$0.15 R$_{L1}$. A possible explanation for the He\,II bright spot is that it is   
formed by overflowing material that reimpacts onto the inner disc 
regions  closer to the primary star. The remarkable  tendency of emission lines to 
fade around superior conjunction of the secondary  is difficult to explain. In Sect.\,4.1 we
will present further details on this behaviour.

\section{Results from photometry}

\subsection{The variability during the persistent state}

\begin{figure*}
\begin{tabular}{c}
\includegraphics[width=2.3in, angle=0]{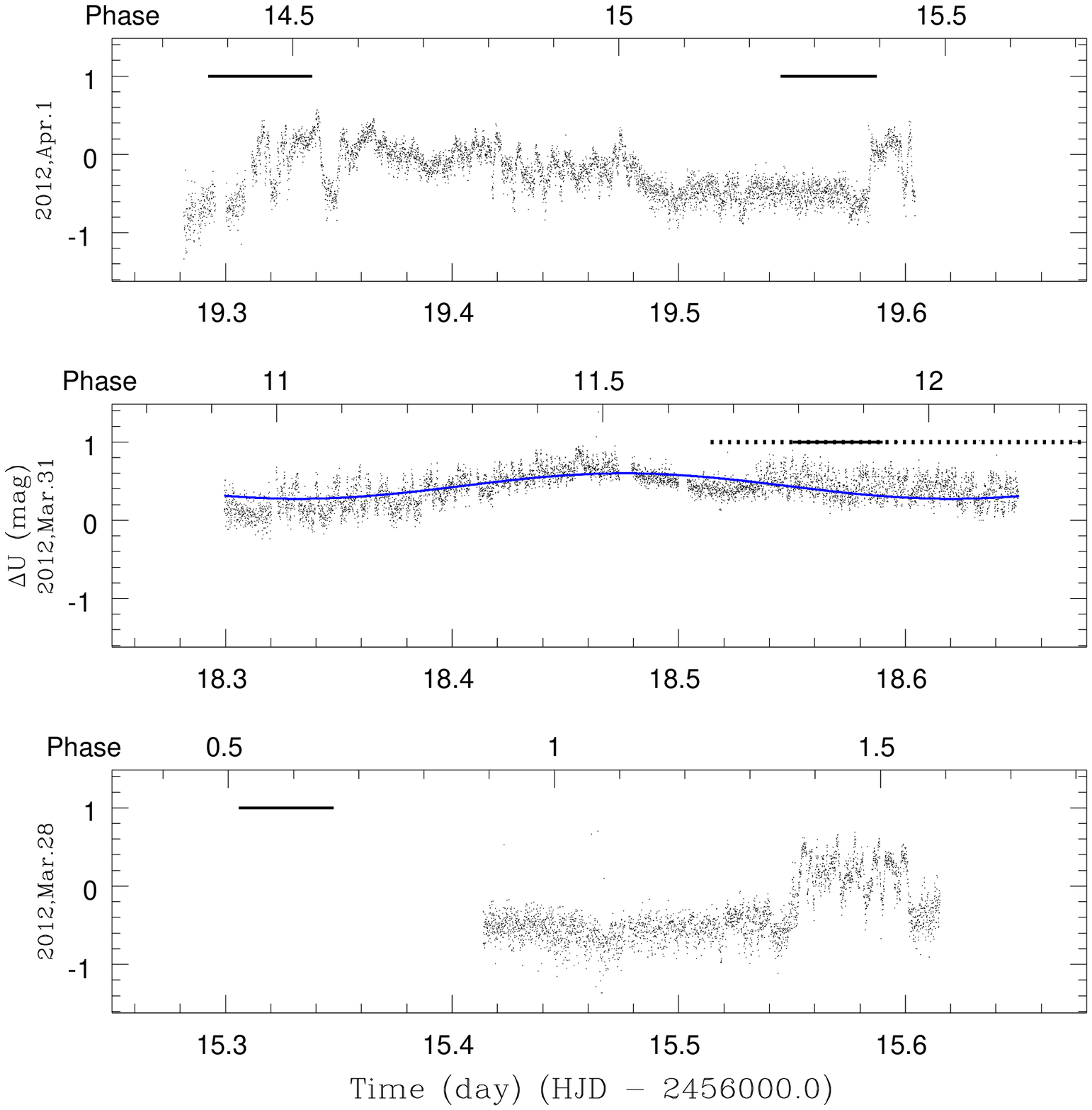}
\includegraphics[width=2.3in, angle=0]{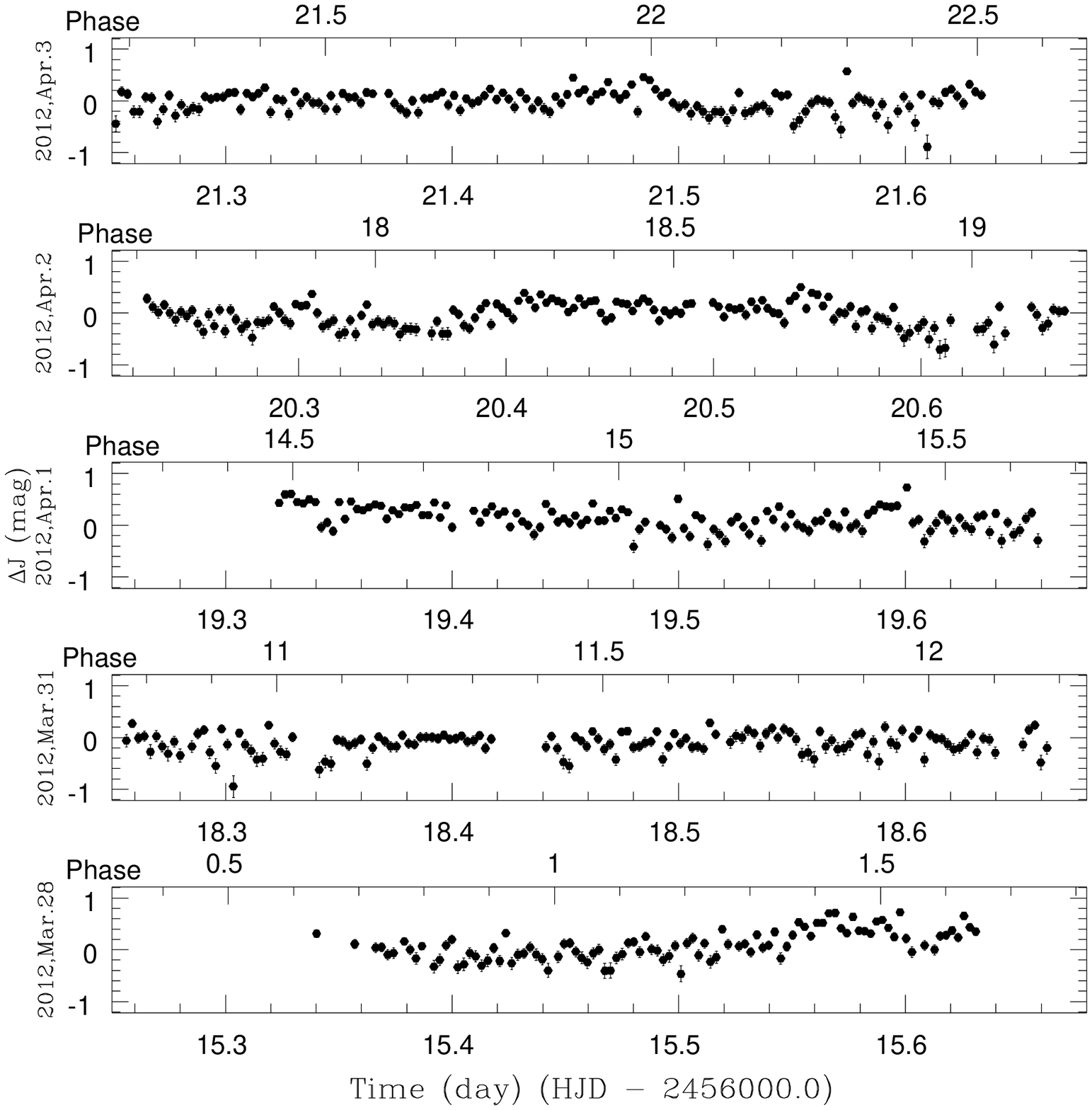}
\includegraphics[width=2.3in, angle=0]{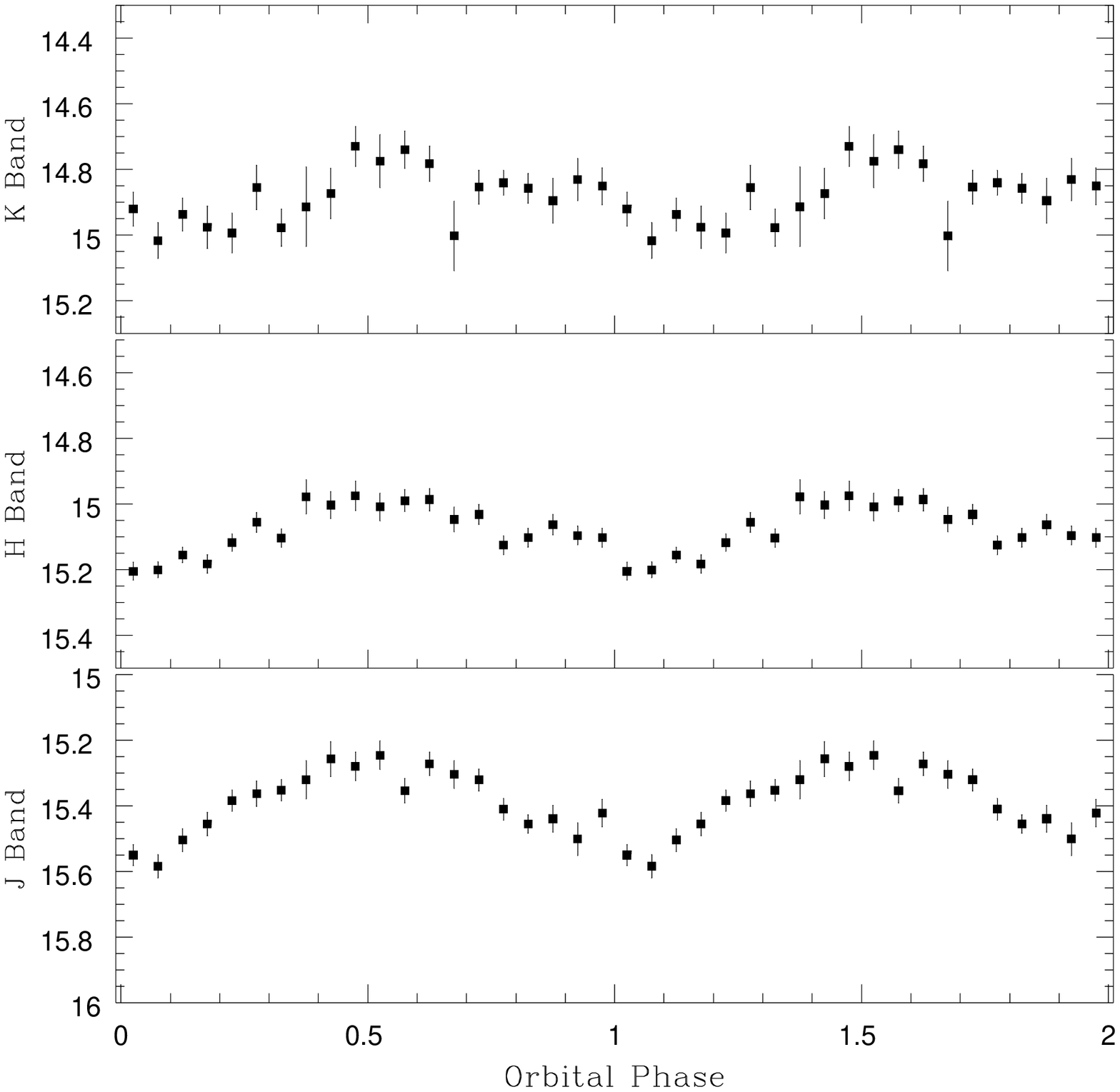}
\end{tabular}
 \caption{{\it Left:} The U-band photometry acquired at the 1.9m SAAO 
telescope during 2012, Mar.-Apr.
showing flares and dips. The spectroscopic coverages with SALT and NTT are also reported as black
thick and dotted lines, respectively.  Orbital phases are reported in the 
upper abscissas. A sine wave function (blue line) at the orbital period is also superimposed
on the U band photometry of Mar. 31. {\it Middle:} The 
simultaneous J-band photometry acquired at the IRSF SAAO telescope showing similar behaviour. {\it Right:}
The nIR light curves folded at the orbital period after removing flares.}
\label{photom2012}
\end{figure*}

The simultaneous optical U band and nIR photometry acquired in 2012 confirms the peculiar behaviour  
detected by 
\citet{Pretorius09}, dM10, \citet{Saitou11} and dM13, with dips, flares and flickering 
occurring in all the filters (Fig.\,\ref{photom2012}). A large amplitude (up to $\sim$ 1\,mag) 
variability is observed without an apparent orbital dependence.
 From Fig.\,\ref{photom2012} the source appears always brighter 
around orbital phases 0.4-0.5. Also, though not recurrent, flares tend 
to occur in this phase range. Similar behaviour is 
observed in subsequent white light photometry  (Kotze, private 
communication). The light curve of Mar. 31 (middle left panel of 
Fig.\,\ref{photom2012}) 
does not show flares for more than one cycle but 
flickering overimposed on a roughly 
 sinusoidal modulation. A formal fit at the fixed orbital period (see 
Fig.\,\ref{photom2012}) gives a time of 
maximum $\rm T_o$ = 2456018.47(2)  and a full amplitude $\Delta U$ = 0.32(4). 

\noindent While the amplitude of the modulation is similar to those observed in
other LMXBs, the light curve is not double-humped as 
it would be expected from 
ellipsoidal  variations of the secondary star 
\citep[]{Shahbaz03}. Single humped light curves with a minimum 
at inferior conjunction of the secondary star instead indicate that the donor 
irradiation is an important source of variability  as also observed during
low states in the MSP binaries SAX\,J1808.42-3658 
\citep[]{Homer01}, PSR\,J102347+0038 \citep[]{Thorstensen05} or 
PSR\,J2215+5135  \citep[]{Breton13,Schroeder14}. 

\noindent We compare the portion of the light curve acquired on Mar. 31 
with the simultaneous SALT and NTT spectra that cover 
cycles 11.7-12.14  (Fig.\,\ref{photom2012}, middle left panel). 
Here an  intense flickering is observed and the 
spectra show strong emission lines, though with different 
shapes. These are signatures of a contribution from the accretion disc, which 
is not expected to produce 
substantial orbital variability \citep[][]{Bayless11}.

\noindent On the other hand, during flares (as those observed on Apr.\,1)  
(Fig.\,\ref{photom2012}, upper left panel) 
the line profiles are complex with absorption and emission 
components with variable shapes. The pre-flares 
and dips are  characterised only by absorptions. Here we recall that 
the pre-flare dips as well as the erratic 
dips seen in X-rays are due to occultation rather than absorption (dM10, dM13).
This peculiar variability cannot be easily reconciled with only  
stationary orbitally locked emitting  regions. A variable extra source of 
light is needed to account by for 
the flares as well as an opaque and localised material that crosses the line 
of sight at the primary inferior conjunction is needed to explain 
the overimposed dips.

\noindent The nIR IRSF photometry
gives the source at: J=15.39(8), H=15.08(8) and K$\rm _s$=14.9(2) 
consistent with the measures 
in 2009 reported by dM10 and by \citet{Saitou11}.   The light curves, though with lower temporal 
resolution, display the same variability seen in the U band (Fig.\,\ref{photom2012}, middle panels).  
Removing the few flares, J-H and 
H-K$\rm _s$ colours indicate bluer spectrum when it brightens. The H,J,$\rm K_s$ 
phase folded light curves (Fig.\,\ref{photom2012}, right panels) reveal a quasi-sinusoidal modulation
with a maximum centred at orbital phase 0.5 and  full amplitudes  
$\Delta J \sim$ 0.4\,mag and  $\Delta H \sim$  0.2\,mag 
(the $\rm K_s$ photometry being too noisy). 
 Correction of the average J and H magnitudes for interstellar reddening 
with E$_{B-V}$=0.11 (dM10), we obtain (J-H)$\rm _o$ = 0.27(11) 
consistent with a late F - early G spectral type \citep{Bilir2008,Straizys2009}. 
At the minimum of the  modulation we obtain (J-H)$\rm _o$ = 0.33(7) corresponding to a 
G5-6\,V spectral type.
 This suggests that the nIR light mainly arises from 
the donor star, as also indicated by the analysis of the SED  (dM13), that
is affected by irradiation. 

\subsection{The photometric variability during the low state}

\begin{figure}
%\begin{tabular}{c}
\includegraphics[width=3.2in, angle=0]{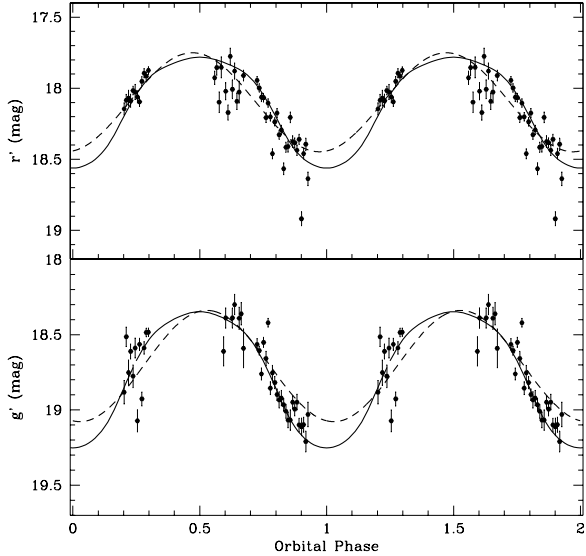}
%\end{tabular}
 \caption{The g' (bottom) and r' (top) light curves in 2013 Dec. folded at the orbital period 
together with a sine wave function (dashed line) and with the simulated light curves obtained with {\sc 
Nightfall} code for $q$=0.3 and $i=69^o$ (solid line) described
 in the text.}
\label{photom2013}
\end{figure}

The REM/ROSS photometry acquired in 2013 gives the source at 
g' = 18.82(5)  and r'=18.19(5), hence
$\sim$2\,mag fainter than in 2009 and 2012. The simultaneity with the NTT spectra of 
Dec. 14 and 15 gives
 further support that XSS\,J1227 has faded to the lowest level ever observed. 
The  light curves, though of poor quality,  reveal a variability consistent with the 
orbital period  (Fig.\,\ref{photom2013}). A sinusoidal fit at the fixed period 
(Table\,\ref{orbparam}) gives full amplitudes $\Delta$ g' = 0.74(3) 
and $\Delta$ r' = 0.70(4). 
%The average colour g' - r' = +0.63(7) allows to obtain \citep{Fukugita96}  B= 19.29 and 
%V=18.45  and hence B-V=0.84. Using the reddening, E$_{B-V}$=0.11 (dM10), it corresponds 
%to (B-V)$\rm _o$=0.73, typical of a G5-G8\,V star \citep{Bilir2008}. This further confirms 
%that the optical light in 2013  arises from the donor star and not from accretion. 
 The large amplitude variation is not easy to explain in terms of the sole contribution 
from the unheated donor star.

\noindent 
 To check if the observed g' and r' light curves can be explained by
an irradiated donor star we simulated both light curves at the refined orbital period using
the {\sc Nightfall} code\footnote{ The {\sc Nightfall} code is available
at http://www.hs.uni-hamburg.de/DE/Ins/Per/Wichmann/Nightfall.html}. 
 We adopt a blackbody temperature of 5500\,K for the unheated donor star assumed to fill its Roche lobe 
and the maximum allowed  temperature (500\,000\,K) for the irradiating star treated as a point 
source.
Given the low quality and poor orbital coverage of our photometric data especially at the inferior and
superior conjunctions of the donor star, we also include in the analysis the secondary star 
radial velocity curve acquired in 2012, which has a denser orbital coverage than that obtained in 
Dec. 2013. We left the mass ratio  $q$ to vary between 0.2 and 
0.4 (see also sect.\,5.1) and obtain $q$=0.29 and a binary inclination $i=69^o$ ( $\rm \chi^2_{red}$=8.9, d.o.f=160). 
The companion star is found to be heated up to 6100\,K. 
The two parameters result to be unconstrained: $q\lesssim 0.5$  and $i\gtrsim 43^o$ at 
3$\sigma$  confidence level. 
Given the low fit quality, we do not attempt to further interpret the reults of the fitting. 
%We also allowed to vary the donor filling factor obtaining a lower limit of 70$\%$.  
The simulated light curves are also shown in Fig.\,\ref{photom2013}.

%% VEDI ROBINSON E.L. 2014 per la curva luce LMXB
%% Bayless et al. 2011 ApJ 730 43
%G5 sptype - Mv=5.1mag 

\section{Discussion}

 Our optical spectroscopy and photometry during the persistent state that characterised 
XSS\,J1227 for about a 
decade have shown a peculiar variability of both emission 
lines and continuum that display dramatic 
quasi-erratic changes  superimposed on a periodic modulation. From the detection of absorption lines 
from the donor  star we have derived orbital period of 6.91246(5)\,h and the first 
spectroscopic ephemeris for this system. 
The dim state into which  XSS\,J1227 entered since the end of 2012 is characterised by the absence 
of  emission lines and the lowest optical flux ever observed, indicating that mass transfer has 
completely switched off and that the donor star dominates the optical light. In the following 
we discuss the  implications of our study  of radial velocities and of optical and nIR light curves.

\subsection{The binary masses and distance}

From absorption lines radial velocity curve of the donor star we have inferred an 
amplitude\footnote{The K$_2$ amplitude could be 
affected by irradiation effects (see \citet{Wade_Horne88}) but, since  
quenching of  absorption lines is not detected, we do not apply the so-called K-correction} K$_2$ = 
261(5)\,km\,s$^{-1}$,  yielding a 
mass function for a circular orbit:

\begin{center}
$ \rm  f(M_1) = M_1 \, sin^3\,i/(1+q)^2 = P_{orb} K^3_2 / 2\,\pi\,G$ = 0.53$\pm$0.03\,M$_{\odot}$. \\
\end{center}

%and its orbital motion from which is heavily affected by irradiation
%(F5\,V spectral type at superior conjunction 
%The absorption lines detected in the optical spectra during the low state indicate a donor star of 
%spectral type G5\,V. The star is affected by X-ray irradiation during the long persistent state, 
%showing changes from ~G5\,V to ~F5\,V  at orbital phases 0. and 0.5, respectively. 

\noindent We also derived the secondary rotational broadening: $V\,sin\,i$=86$\pm$20\,km\,s$^{-1}$.
Since  $V\,sin\,i$ scales with  K$_2$ and it is a function of the mass 
ratio $q$ \citep{Horne86}, we estimate $q$ = 0.25$\pm$0.12. 

\begin{figure}
%\begin{tabular}{c}
\includegraphics[width=3.2in, angle=0]{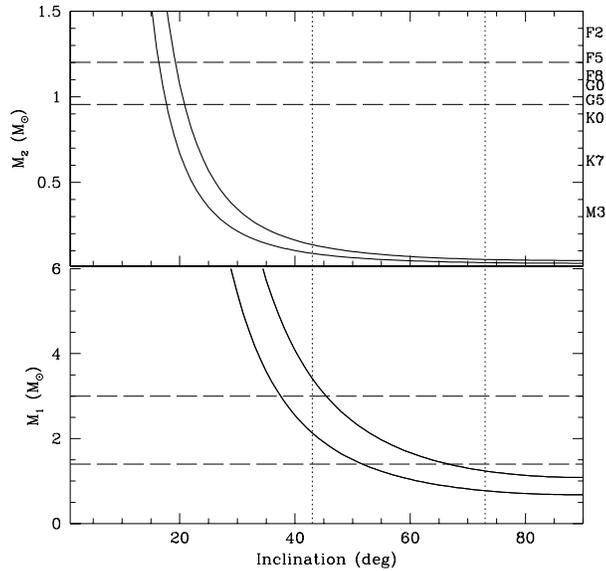}
%\end{tabular}
\caption{The primary (bottom) and secondary (top) masses versus the binary inclination angle $i$  adopting a mass
ratio  $q$=0.13-0.43 and mass functions f(M$_1$)=0.53\,M$_{\odot}$ (bottom) and  f(M$_2$)= 
0.02\,M$_{\odot}$ (top). The binary inclination range $43^o \le i  \le 73^o$ (see text) is shown as
dotted vertical lines. The NS  mass  values $\rm M_1$=1.4\,M$_{\odot}$ and 3\,M$_{\odot}$ are shown 
as long dashed horizontal lines.
The  masses for donors between spectral type G6\,V and F5\, are  also reported as horizontal long dashed lines.
}
\label{mass_incl}
\end{figure}

\noindent From the radial velocity variations of the wings of H$_{\beta}$ emission line 
we   estimated  an  amplitude of the compact object orbital motion  
K$_1$ =89(23)\,km\,$^{-1}$, although this should be regarded as an upper limit (see e.g. 
\cite{Marsh88}). This yields a mass function $\rm f(M_2) = 0.02\,M_{\odot}$.  Using
the ratio of the radial velocity amplitudes $\rm K_1$ and $\rm K_2$ we obtain a mass ratio 
$q$=K$_1$/K$_2$ = M$_2$/M$_1$ = 0.34$\pm$0.09 that is within 1$\sigma$ consistent 
with that derived above.  We then conservatively adopt a mass ratio $q$=0.13-0.43,
the above values for $\rm f(M_1)$ and $\rm f(M_2)$ to infer the allowed ranges of the 
binary component masses. We also make use of the binary inclination limits $i \gtrsim 43^o$ 
(Sect. 4.2) and $i \lesssim 73^o$ due to the absence of X-ray eclipses. In 
Fig.\,\ref{mass_incl} (lower panel), for a  canonical NS mass 1.4\,M$_{\odot}$, the binary  inclination is limited to
$ 50^o \lesssim i \lesssim 65^o$. Allowing a wider range of NS masses 
$\rm M_1 = 1.4-3\,M_{\odot}$, which is a likely possibility because fastest spinning NS tend 
to be massive  \citep{Roberts13}, the binary inclination range is: $ 45^o \lesssim i \lesssim 65^o$. 
Within these values the secondary mass is limited to 
 $\rm M_2 \simeq$ 0.06-0.12\,M$_{\odot}$, implying an extremely undermassive
donor  for its spectral type (Fig.\,\ref{mass_incl}, upper panel). 
We then adopt the latter ranges for the donor star mass and for the binary inclination.

\noindent Extremely low mass donors are indeed found in many 
MSP binaries with short ($\rm P_{orb} \ll $ 24\,h) 
orbital periods,  which are eroded by strong irradiation
\citep{Chen13,Roberts13}. They are broadly divided into the 
so-called black widows, harbouring strongly 
ablated and likely degenerate secondaries   with  $M_2 \ll 0.1\,M_{\odot}$ (typically 
0.02-0.04\,M$_{\odot}$) 
and the so-called  redbacks which have non-degenerate secondaries with masses  $\rm M_2 \simeq 0.1 - 
0.4\,M_{\odot}$ \citep{Roberts13}.  The donor in XSS\,J1227 then appears to be a new member of the 
redback group, that now would count with seven systems. Higher resolution spectroscopy and 
better photometric coverage are  needed to constrain the mass ratio and orbital inclination.

\noindent Previous estimates of 
the distance to XSS\,J1227 (1.4 - 3.6\,kpc) were based on high state nIR photometry 
(dM10,dM13).  The optical magnitudes during the dim state, ascribed to the donor star, 
allow us to determine a new  distance.  The average and lowest observed 
g' and r' magnitudes corrected 
for extinction ($\rm E_{B-V}$=0.11), when translated into the V band, give V=18.11 and 
V=18.45, respectively. 
 The latter, if due to a G5\,V star ($\rm  M_V$=5.1), would imply a distance of 3.4 -- 
4.7\,kpc. However, a G5 main sequence star has a larger radius than the donor in XSS\,J1227. 
Assuming
the secondary fills its Roche lobe and using $\rm M_2 = 0.06 - 0.12\,M_{\odot}$ and the binary orbital period, 
we obtain $\rm R_2 = 0.34 - 0.42\,R_{\odot}$, which  gives $\rm M_V \sim 6.9-7.2$\,mag and then
a distance of 1.8-2.0\,kpc.

\subsection{Irradiation effects }

We investigate the effect of irradiation 
due to the high  energy emission from the compact star.  Irradiation increases 
the local effective temperature of the donor at the inner face.  
The effect of an anisotropic irradiation
is such that the energy outflow through the surface layers of the companion is blocked over 
a fraction of its surface \citep{Ritter+00}. Consequently the heated face of the donor star
emits a fraction of the incident flux: $\rm F_{irr} = W\,F_{inc}$, where $\rm F_{irr} = 
\sigma\,T_{irr}^4$ is the flux of the irradiated surface of the donor,  $\rm F_{inc}$ is the 
component normal to  the stellar surface of the incident flux and $\rm W$ is
the reflection albedo, which, for stars in the temperature range 2000-10000\,K, is between 
0.5 and 1.0 \citep{Claret01} for atmospheres that are convective and in radiative equilibrium, 
respectively. Following \citet{Ritter+00}, $\rm F_{inc} = \eta\, 
(L_{high,emiss.}/ 4\,\pi\,a^2)\,  h(\theta)$, with $\eta$ an efficiency factor that is unity if 
the  high energy luminosity is radiated isotropically. XSS\,J12270 has been found in a prolonged
persistent state until late 2012. We then adopt  $\rm 
\eta  L_{high,emiss.}  = L_{X,\gamma} \sim 5 \times 10^{34}\,  erg\,s^{-1}$, using the 
bolometric X-ray/gamma-ray fluxes derived in dM10 and dM13 and d=1.9\,kpc (Sect.\,5.1), 
a binary orbital separation $\rm a \sim 2.2\,R_{\odot}$, for 
q=0.13-0.43 and $i$ = 45$^o$-65$^o$, and $\rm h(\theta) \simeq cos(\theta)$ 
(see \citet{Ritter+00} for details). For a typical incidence angle $\theta =45^o$, 
we derive  $\rm T_{irr} \sim$ 5800-6800\,K, for albedos $\rm W$ in 
the range 0.5-1.0, respectively. Thus the heating of the secondary star inferred from the optical
spectra is  broadly  consistent with irradiation by the high energy X-rays and gamma-rays 
emitted during the persistent state. 
Furthermore, \citet{Papitto14} modeled 
the X-ray and gamma-ray emissions in a propeller configuration with a total luminosity $\rm  
L_{rad} \sim 1.5\times 10^{35}\,  erg\,s^{-1}$. This would imply an efficiency $\eta \sim$ 0.3.
On the other hand, the current low state into which XSS\,J1227 has entered is characterised by a
a decrease in the X-ray  flux  by  a factor of $\sim$30 \citep{Bassa13,Bassa14,Bogdanov14} 
also accompained by a dimming of the gamma-ray \emph{Fermi}/LAT flux \citep{Tam13}. 
Notwidthstanding this, the 
secondary star is not expected  to cool substantially, because the thermal relaxation 
timescale is 
of the order of the thermal timescale of the convective envelope ($\tau_{CE} \simeq \tau_{KH} \propto \rm 
(RL)^{-1}$) \citep{Ritter00}. 
Hence, while the optical  light during the persistent state in 2012 could be heavily contaminated 
by  the accretion disc, the modulation during the low state is dominated by the irradiated donor.
In Sect.\,4.2 we showed the {\sc Nightfall} simulation for q=0.29 and i=69$^o$ and a heated 
secondary up to 6100\,K that reproduces the observed large amplitude $\rm \Delta M \sim$0.7
orbital modulation in Dec. 2013. It also predicts amplitudes 
$\rm \Delta M \sim$0.3-0.4  in J,H and K bands consistent with the observed nIR modulation observed
in 2012, suggesting that the donor dominates at these wavelengths  even in the higher state.

\noindent  Single humped orbital modulations are observed in many black widows (up to 
$\sim$2\,mag) as well as in redbacks ($\sim$0.2-0.8\,mag) when in low states 
\citep{Thorstensen05,Archibald13,Breton13,Schroeder14}.  
The orbital modulation is caused by a combination of irradiation, which produces a maximum at the 
superior conjunction of the companion, and ellipsoidal variations, which yield maxima at 
quadrature. Single peaked  optical modulations are suggestive of negligeable contribution of 
ellipsoidal variations and the amplitudes depend on the binary inclination.   
Examples are the redbacks PSR\,J2215+5135 and PSR\,J102347+0038, whose 
optical orbital modulations have amplitudes of $\sim$0.8\,mag and $\sim$0.4\,mag respectively. These
 were modeled with  mildly irradiated donors, filling their Roche lobe, and viewed at moderate 
inclination angles 
$\sim$48-60$^o$ \citep{Schroeder14}  or $\sim 70^o$ \citep{Breton13} and
$\sim$40$^o$ \citep{Thorstensen05}, respectively.  
We conclude that XSS\,J1227 is a redback with a moderate irradiated donor.

\subsection{A low-mass X-ray binary in a propeller state}

 Drawing analogy with PSR\,J102347+0038, \citet[][]{Papitto14} proposed XSS\,J1227 to be a LMXB 
in a  propeller state where a rapidly spinning MSP inhibits most of the mass transferred 
from the donor star to accrete. In this 
configuration matter forms a shock at the magnetospheric boundary where 
the high energy gamma-ray emission as well as most of the X-ray radiation take place. 
The recent discovery of  1.69\,ms radio pulsations \citep{Roy14} gives support to this hypothesis. 

\noindent From our spectroscopic study during the persistent state there is evidence that 
an accretion disc is present. The Doppler map of  H$_{\beta}$ emission reveals a spot located 
at  $\sim$0.55 R$_{L1}$, that for $q$=0.25 translates 
into $\sim 6\times 10^{10}$\,cm. This provides an estimate of the outer disc radius. On the other 
hand, the He\,II
emission spot is localised at $\sim$0.15\,$\rm R_{L1}$ that translates into $\sim 
1.6\times 10^{10}$\,cm. A consistent value is obtained from H$_{\beta}$ emission line wings 
(FWZI/2$\sim$  1000\,km\,s$^{-1}$) assuming they trace  the keplerian velocity of the 
innermost disc regions contributing to this line ($\sim 1.9\times 10^{10}$\,cm). This value is
also consistent with that of a hot emitting region  ($\rm T_{hot} \sim 13000$\,K)  
inferred from the analysis of the SED (dM13). 
 
For a pulsar rotating at a period of 1.69\,ms the light cylinder radius 
at which the corotation velocity exceeds the speed of light is
$\rm R_{LC} = c\,P/2\,\pi \sim$ 80\,km. 
At this radius the pulsar emission prevents the plasma to reach the 
surface and a rotation-powered pulsar would turn on. If the magnetosphere is squeezed
below $\rm R_{LC}$ at distances of the order of the  corotation radius   $\rm R_{co} = 
(G\,M_{NS}\,P^2 / 4\,\pi)^{1/3} \sim$ 24\,km the radio pulsar would turn off and accretion take place. 
For XSS\,J1227 during the persistent higher state, \citet[][]{Papitto14} estimate an inner disc 
radius of $\sim$40\,km  at which  the interface magnetic field strength is limited
to 3-6\,MG. At this radius the  disc is dead and the propeller process 
takes place where  the X-ray and  gamma-ray emissions originate. 
Then the disc regions contributing to the optical emission lines and  
UV light (dM13) are clearly located further out. The tight correlation of
X-ray and UV flares further implies that these regions are reprocessing sites of 
the X-ray emission.

\noindent 
Although the propeller model does not necessarly require material to be expelled by the 
system (it can fall back and rebuild the disc) \citep{Papitto14}, it is conceivable that matter is lifted 
off
from the orbital plane.  The tendency of flare/dip pairs to occur at superior conjunction of
the donor star, the fading of emission lines in  this phase range and the night-to-night variability
of this behaviour could be related to the propellor state of the MSP, in which material is propelled
within  the Roche lobe of the primary and intercepts the line of sight.
The remarkable flaring 
and  dipping behaviour in  XSS\,J1227 seems now to be also present in other MSPs. 
PSR\,J1023+0038,  which recovered accretion around mid 2013, also displays an X-ray light 
curve with dips  and a strong variability \citep{Patruno14}.

\noindent XSS\,J1227 has entered in the faintest state ever observed since the end of 2012 
characterised by absence of accretion tracers in the spectra. 
The decay in the X-ray flux observed by \emph{Swift}/XRT \citep{Bassa14} 
and \emph{XMM-Newton} \citep{Bogdanov14}  
and the possible decrease in the \emph{Fermi}/LAT 
countrate \citep{Roy14} further confirm accretion has switched off. 
XSS\,J1227  has therefore transited from a LMXB to a MSP phase, the 
opposite way of PSR\,J1023+0038 that has recently brightened in the X-rays,  optical and in 
the high energy gamma-rays \citep{Stappers14,Takata14,Patruno14}.

\noindent  In order to better constrain orbital parameters and energetics of XSS\,J1227
multi-wavelength follow-ups in the ongoing unprecedented faint 
state of this intriguing source are needed.

\section*{Acknowledgments}
ESO/NTT observations were acquired under programmes 088.D-0311 and 092.D-0588.
Some of the observations reported in this paper were acquired at the
Southern African Large Telescope (SALT) under programme
2011-3-RSA OTH-025.  
The ESO REM observations were obtained under programme DDT-REM:28903.
DdM wishes to thank the REM team to promptly schedule the observations. 
EM thanks ESO La Silla support astronomer and the hospitality at INAF $-$ Astronomical 
Observatory of Capodimonte. 
JC thanks Dr. P. D'Avanzo for his help in acquiring the 2013 
ESO/NTT spectra. We are grateful to Dr. Nagayama and Dr. D. Foster for their
support in IRSF/SIRIUS observations and data reduction. 
JC acknowledges support by the Spanish MINECO grant 
AYA2010-18080. 
 We acknowledge the use of the {\sc Nightfall} program for the light
curve synthesis of eclipsing binaries written by Rainer Wichmann. We thank
the anonymous referee for useful suggestions to improve the manuscript.
 
\bibliographystyle{mn2e}
\bibliography{biblio}

\vfill\eject

\end{document}